\documentclass[prb, twocolumn]{revtex4}
\usepackage{graphicx}
\usepackage{dcolumn}
\usepackage{bm}

\begin{document}

\title{Structure of the icosahedral Ti-Zr-Ni quasicrystal}
\author{R. G. Hennig}
\email[]{rhennig@mps.ohio-state.edu}
\homepage{http://www.physics.ohio-state.edu/~rhennig}
\affiliation{Department of Physics, Ohio State University, Columbus, OH 43210, USA}
\author{K. F. Kelton}
\author{A. E. Carlsson}
\affiliation{Department of Physics, Washington University, St. Louis, MO
  63130, USA}
\author{C. L. Henley}
\affiliation{Laboratory of Atomic and Solid State Physics, Cornell University,
  Ithaca, NY 14853, USA}

\date{\today}

\begin{abstract}
  The atomic structure of the icosahedral Ti-Zr-Ni quasicrystal is
  determined by invoking similarities to periodic crystalline phases,
  diffraction data and the results from {\it ab initio} calculations.
  The structure is modeled by decorations of the canonical cell tiling
  geometry. The initial decoration model is based on the structure of
  the Frank-Kasper phase $W$-TiZrNi, the 1/1 approximant structure of
  the quasicrystal. The decoration model is optimized using a new
  method of structural analysis combining a least-squares refinement
  of diffraction data with results from {\it ab initio} calculations.
  The resulting structural model of icosahedral Ti-Zr-Ni is
  interpreted as a simple decoration rule and structural details are
  discussed.
\end{abstract}



\maketitle

\section{Introduction}

Quasicrystals are solids that combine long-range translational order
with rotational symmetries that forbid {\em periodic} translational
symmetry. To understand the physical properties of quasicrystals, and
in particular, the conditions under which nature prefers quasiperiodic
to periodic order, it is necessary to determine their atomic
structure. The aperiodic nature of quasicrystals requires new methods
for structural analysis. In spite of significant progress in recent
years, the complete atomic structure of quasicrystals still remains an
unsettled question.~\cite{Boudard99}

The purpose of this paper is to determine the atomic structure of the
thermodynamically stable icosahedral TiZrNi quasicrystal ($i$-TiZrNi).
Although this material is the best-ordered among the known Ti-based
quasicrystals, only polycrystalline samples with small grain sizes are
available, which limits the information available from diffraction
experiments. Therefore this paper develops a new method for structural
analysis, based on the real-space structure description, whereby
(several kinds of) tiles fill space and each kind of tile receives an
identical ``decoration'' by atomic species, just as unit cells do in
ordinary crystallography.

The known quasicrystal structures fall into two main
types.~\cite{Hen86a} In a Frank-Kasper type structure, neighboring
atoms always form tetrahedra, and the constituent species divide into
``small'' atoms (with icosahedral coordination) or ``large'' atoms
(with coordination number up to 16).  In an Al-transition metal
(Al-TM) structure, the transition metals are commonly surrounded by an
icosahedron of Al, and Al atoms often form octahedra.  In either type,
icosahedral quasicrystals (and closely similar crystalline phases
called ``approximants'') may be built from large ($\sim$50 atom)
clusters with icosahedral symmetry: the ``Bergman
cluster''~\cite{Bergman57} in the Frank-Kasper case, or the ``Mackay
icosahedron''~\cite{Els85} in the Al-TM case.  The Ti-based class of
quasicrystals appears to fall on the borderline between the two main
types~\cite{Hen86a}; thus, though $i$-TiMn and $i$-TiCr fall in the
Al-TM class, our subject $i$-TiZrNi is of the Frank-Kasper type.  In
developing its structural model, we will take advantage of structural
relationships to both the $i$-AlMnSi and the Frank-Kasper
quasicrystals.

\subsection{Methods of Structural Investigation}

Many thermodynamically stable quasicrystals such as
$i$-AlCuFe~\cite{Tsai87}, $i$-AlPdMn~\cite{Tsai90}, and
$d$-AlNiCo~\cite{Ritsch96} within the Al-TM metal class, $i$-AlCuLi
and $i$-MgZnRE (RE=rare earth)~\cite{Niikura94,Tsai94} in the
Frank-Kasper class, as well as the new binary quasicrystal
$i$-CdYb~\cite{Tsai00}, can be grown as single grains up to several
millimeter size, which allows the measurement of single crystal
diffraction data.  (Of those mentioned, all except $i$-AlCuLi also
exhibit long-range order, i.e., instrumentally sharp Bragg peaks.)
Using a variety of procedures~\cite{Steurer99,Boudard99,Elser99},
atomic structures have been refined for many of these quasicrystals,
such as $i$-AlCuLi~\cite{Boissieu91, Yamamoto92, Qiu95},
$i$-AlCuFe~\cite{Cornier91, Katz93, Cockayne93} and
$i$-AlPdMn~\cite{Boudard91, Janot92, Boissieu94}.

All these structures were parametrized as cuts by a three-dimensional
hyperplane through a density in a six-dimensional
hyperspace.~\cite{Bak86,Gratias88,Janot89b} The difficulty with the
six-dimensional approach is that only an average structure can be
determined and disorder in the quasicrystal can not be treated
appropriately. This, combined with experimental limitations such as
truncation effects, usually produces a small number of spurious atoms
with unphysical coordination numbers or small interatomic separations
in structural models. A purely crystallographic approach alone might,
therefore, not be able to fix all the positions and chemical
identities of the atoms in quasicrystalline structures.  It is
furthermore infeasible for quasicrystalline alloys with small grain
sizes for which only powder diffraction data are available.

Additional experimental and theoretical data, such as structural
information from related approximant phases and results from total
energy calculations, can be helpful in the development of realistic
structure models for quasicrystals. To incorporate this information a
real space approach is desirable.~\cite{Henley00,Mih96a} A common real
space approach to the structure of quasicrystals is the tiling
decoration.  One splits the structural analysis into two parts: (1)
the problem of the determination of the correct tiling geometry, and
(2) the atomic decoration of the tiles. An advantage of the decoration
viewpoint is different calculations ({\it ab initio} LDA,
pair-potentials, diffraction simulation) may be performed using
periodic tilings ("approximants") with larger or smaller unit cells,
depending on the calculation's computational burden, and the results
can be compared or combined in a systematic fashion.

The first tiling-decoration approaches were based on similarities to
periodic crystalline structures.~\cite{Els85,Hen86a} In later work
energetic considerations were taken into account. Examples of such
investigations are the work by Kraj{\v c}i and Hafner who investigated
the structure of the $i$-AlMgZn quasicrystal using a decoration of the
three-dimensional Penrose tiling and pair potentials based on the
pseudopotential theory.~\cite{Kra92} Windisch et al.\ studied the
atomic and electronic structure and the lattice dynamics of $i$-AlCuLi
using molecular dynamics simulations and density-functional
calculations of decorated Penrose and canonical cell
tilings.~\cite{Windisch94a, Windisch94b} The structure of the
(metastable) $i$-AlMn quasicrystal has been studied by Mihalkovi{\v c}
et al.\ using canonical cell tiling decorations and
pairpotentials.~\cite{Mih96a, Mih96b} Recently Mihalkovi{\v c} et al.
investigated the structure of decagonal AlCoNi using Monte-Carlo
calculations with tile updates.~\cite{Mih01}

All these investigations are based on the energetics of structural
models and do not incorporate diffraction data into the modeling
process. To derive a reliable structural model of a quasicrystal, a
combination of the crystallographic approach based on diffraction data
with information from total energy calculations and relaxation
studies is highly desirable. In this paper we present the first study
combining a least-squares refinement of X-ray and neutron diffraction
data with structural information obtained from {\it ab initio}
relaxations for approximant structures.

\subsection{The Ti-Zr-Ni quasicrystal}

Few studies have been performed for quasicrystalline alloys with grain
sizes so small that only powder diffraction data are available. Early
work was devoted to the metastable system
$i$-AlMnSi.~\cite{Cahn88,Janot89a,Duneau89} Another class of such
quasicrystals are the Ti-based icosahedral alloys. As early as 1985,
Zhang et al.~\cite{Zha85} discovered a metastable icosahedral phase in
Ti-Ni-V.  In the following years, further icosahedral phases in
Ti-based alloys containing other 3d transition metals such as V, Cr,
Mn, Fe, Co and Ni were
found.~\cite{Don86,Kel88,Dun88,Hol89,Kel90,Zha91,Lib95} However, no
stable quasicrystalline phase of this class has been discovered yet.
Investigations of projection-based models of the structure of the
Ti-Mn quasicrystal by Phillips et al.\ using tight-binding recursion
and pair-potential based energy
calculations~\cite{Phillips91,Phillips92} showed that the Mackay
clusters form the backbone of the structure. Furthermore small
additions of Si and O, which are crucial for the formation of the
Ti-Cr and Ti-Mn quasicrystal~\cite{Zha90,Lib95}, were found to
energetically stabilize the structure.~\cite{Hennig97}

In this work the experimentally well-studied icosahedral Ti-Zr-Ni
quasicrystal is considered. As early as 1990, Molokanov and
Chebotnikov~\cite{Mol90} discovered a metastable icosahedral phase in
Ti-Zr-Ni and in 1997 a thermodynamically stable icosahedral Ti-Zr-Ni
phase was discovered by Kelton et al.~\cite{Kelton97} The $i$-TiZrNi
quasicrystal is formed by either rapid quenching or solid state
transformation at temperatures of $500^\mathrm{o}$C to
$600^\mathrm{o}$C, generally leading to a fine microstructure of
quasicrystal and crystal phases. The icosahedral phase is rather
disordered with a coherence length of about 350\,\AA, resulting in a
small number of measurable structure factors.  Therefore, an atomic
structure determination by diffraction experiments alone is
impossible. Consequently little is known about the atomic structure of
$i$-TiZrNi.

There is hope of obtaining information on the atomic arrangement in
$i$-TiZrNi by studying the structure of related periodic
``approximants'', meaning crystal structures whose unit cells are
fragments of some quasicrystal. There are several competing periodic
phases known in the Ti-Zr-Ni phase diagram.  Besides the binary phases
$\alpha$-TiZr, $\delta$-Ti$_2$Ni~\cite{Yurko59} and
$l$-Zr$_2$Ni~\cite{Kirkpatrick62}, there are two ternary phases; a
cubic Frank-Kasper type structure ($W$-phase) and a hexagonal Laves
type phase $\lambda$-TiZrNi.~\cite{Molokanov89} The structures of the
$W$-TiZrNi phase, which can be interpreted as a Fibonacci
1/1~approximant of $i$-TiZrNi, and of the $\lambda$-TiZrNi phases were
determined by a Rietveld refinement using X-ray diffraction and
neutron scattering data. Structural details were confirmed by {\it ab
  initio} calculations.~\cite{Kim98,Hennig00,Majzoub01} The $W$-TiZrNi
phase is closely related to the quasicrystal
phase.~\cite{Kim97,Hennig00} They both form at similar compositions
and temperature ranges and grow with a coherent crystallographic
orientation. EXAFS studies indicate a similar local structure in both
phases.~\cite{Sadoc00} The presence of a reversible phase
transformation between the icosahedral phase and the $W$-phase at
about $600^\circ$C indicates that $i$-TiZrNi is a low-temperature
phase, in contrast to most other quasicrystals.~\cite{Dav00} This
opens the exciting possibility that an (aperiodic) $i$-TiZrNi
quasicrystal, or a very similar large-cell crystal, could be a ground
state phase.

\subsection{Contents of this paper}

The goal of our work is to model the ideal structure of the $i$-TiZrNi
quasicrystal. The energetic stability and electronic structure of
approximants to the $i$-TiZrNi quasicrystals, as well as the competing
phases, are the topics of a subsequent paper. In this paper we present
a structural analysis of $i$-TiZrNi based on a new method, combining
powder diffraction data and information from {\it ab initio}
calculations in form of a constrained least-squares analysis. In
Section~\ref{Decoration-Section} the initial decoration model is
derived. We base our structure on the canonical cell tiling geometry,
which is introduced in Section~\ref{CCT-Section}. As our starting
point an initial atomic decoration of the canonical cells is
determined from the atomic structure of the 1/1 approximant $W$-TiZrNi
(Sec.~\ref{Initial-Section}). The following
Section~\ref{Refinement-Section} on the optimization of the decoration
model is the core of the paper. The constrained least-squares method
of refinement is introduced. This approach incorporates structural
information from {\it ab initio} relaxations as constraints into a
least squares refinement of the diffraction data. The resulting
structure can be understood as a simple decoration model. This and
further details of the structural model, which are important for
comparisons to experimental data such as EXAFS measurements, are
discussed in Section~\ref{Sec:Structure}.

\section{First decoration model} \label{Decoration-Section}

In this section we derive the initial decoration model. The underlying
tiling geometry of our model is given by the canonical cell tilings.
The canonical cell tilings provide a tiling model with a high packing
fraction of icosahedral clusters. The location of the atomic sites
decorating the canonical cells in our model are based on structural
features of the $W$-phase.

\subsection{Structure of the $W$-phase} \label{W-Section}

The $W$-TiZrNi phase is a periodic crystalline phase closely related
to the icosahedral quasicrystal. Its composition
Ti$_{50}$Zr$_{35}$Ni$_{15}$ is close to the composition of the
quasicrystal Ti$_{41.5}$Zr$_{41.5}$Ni$_{17}$ and it grows with a
coherent crystallographic orientation with respect to the
quasicrystalline phase. The structure of the $W$-phase was determined
using X-ray diffraction and neutron scattering data and confirmed by
{\it ab initio} calculations.~\cite{Kim98,Hennig00}

It consists of a body centered cubic packing of icosahedral Bergman
clusters connected by a few ``glue'' sites (see Table
\ref{tab:Approximant}). The Bergman cluster as shown in
Figure~\ref{fig:Bergman} is composed of a central Ni atom surrounded
by a small icosahedron of 12 Ti atoms. A second icosahedron is formed
by placing 12 Ni atoms directly outside the vertices of the small
icosahedron. Finally 20 Zr atoms are placed on the faces of the large
icosahedron. The remaining 36 ``glue'' sites consist of 24 Ti atoms,
forming a hexagon that separates the Bergman clusters along the
threefold $\langle 111\rangle$ directions, and 12 Ti and Zr atoms,
forming a rhombus that divides the Bergman clusters along the twofold
$\langle 100\rangle$ directions. The structural analysis and {\it ab
  initio} calculations revealed that chemical disorder in the
$W$-phase is abundant only between the chemically similar Ti and Zr
atoms on the ``glue'' sites but no disorder is found on the cluster
sites.~\cite{Hennig00}

\begin{table}[tb]
  \caption{Crystallographic description of the phase $W$-{\rm
      TiZrNi}~\cite{Hennig00} showing the chemical occupation of the lattice
      sites. The positions of the
      sites are given in units of the lattice constant.}
  \label{tab:Approximant}
  \renewcommand{\tabcolsep}{0.3cm}
  \begin{tabular}{c c c c c c}
    \toprule
    \multicolumn{6}{c}{cI162 \hspace{0.8cm} Space group Im$\bar 3$\hspace{0.8cm}
      $a_o=14.30\,$\AA} \\
    \colrule
    \multicolumn{2}{c}{Site} & $x$ & $y$ & $z$ & Occupancy \\
    \colrule
    1 & $\alpha_1$ & 0     & 0     & 0     & Ni(1.0) \\
    2 & $\beta_1$  & 0     & 0.103 & 0.165 & Ti(1.0) \\
    3 & $\alpha_2$ & 0     & 0.193 & 0.310 & Ni(1.0) \\
    4 & $\gamma_1$ & 0.186 & 0.186 & 0.186 & Zr(1.0) \\
    5 & $\gamma_1$ & 0     & 0.307 & 0.114 & Zr(1.0) \\
    6 & $\delta_1$ & 0.204 & 0     & 0.5   & Zr(0.6), Ti(0.4) \\
    7 & $\delta_2$ & 0.145 & 0.187 & 0.401 & Ti(1.0) \\
    8 & $\beta_2$  & 0.412 & 0     & 0.5   & Zr(0.5), Ti(0.5) \\
    \botrule
  \end{tabular}
\end{table}

\begin{figure}[p]
  \caption{Structure of the Bergman cluster consisting of a central Ni atom (left),
    a small icosahedron of Ti atoms (center) and a large icosahedron of Ni atoms with
    Zr atoms on the face centers (right).}
  \label{fig:Bergman}
\end{figure}

The atomic structure of the $W$-phase is similar to the structure of
the Frank-Kasper phases
$\mathrm{Al_{40}Mg_{40}Zn_{20}}$~\cite{Bergman57} and
$\mathrm{Al_{56}Li_{32}Cu_{12}}$~\cite{Audier88}. The $W$-TiZrNi
structure differs from these in (i) the occupation of the central site
of the Bergman cluster, which is empty in the cases of Al-Mg-Zn and
Al-Li-Cu, and (ii) the chemical disorder, which occurs between the
smaller components Al and Zn or Cu in Al-Mg-Zn and Al-Li-Cu
respectively and between the larger components Ti and Zr in the case
of Ti-Zr-Ni. Furthermore, the analysis for $i$-TiZrNi showed chemical
ordering on the sites of the Bergman cluster.

\subsection{Canonical cell tiling} \label{CCT-Section}

In modeling the quasicrystal structure the canonical-cell tiling (CCT)
model is used.~\cite{Hen91a} The packing rules permit periodic, random
and in principle quasiperiodic tilings, although no rule for a
quasiperiodic tiling has been found yet.~\cite{Hen91a,Mih96a} A large
number of periodic approximants structures, ranging from a 1/1 up to a
31/24 approximant, are known for this tiling. In order to use the
CCT's for a structural description of quasicrystals and for a
refinement of decoration models to experimental diffraction data,
predictions for the structure factors of the icosahedral structure are
needed (see App.~\ref{App:Diffraction} and~\ref{App:Amplitudes}).

Canonical cell tilings consist of four kinds of disjoint cells denoted
$A$, $B$, $C$ and $D$.  Their nodes are joined by two types of
linkages, $b$ and $c$ bonds, which run in twofold and threefold
symmetry directions respectively (see Fig.~\ref{fig:cct}). There are
three different kind of cell faces. The $X$ face has the shape of an
isosceles triangle formed by one $b$ bond and two $c$ bonds, the $Y$
face with a shape of an equilateral triangle is formed by three $c$
bonds and finally the rectangular $Z$ face is formed by two $b$ bonds
and two $c$ bonds.
\begin{figure}[p]
  \caption{The four canonical cells. The b-bonds and c-bonds are represented by the
    double and single lines respectively. The face names are also
    indicated.}
  \label{fig:cct}
\end{figure}

\begin{figure}[p]
  \caption{Decomposition of the canonical cells tiling into prolate and
    oblate rhombohedra as well as the rhombic dodecahedron.}
  \label{fig:Decompose}
\end{figure}
The simplest quasiperiodic tiling with icosahedral symmetry is the
Penrose tiling. Its tiles are the prolate rhombohedron ($PR$) and
oblate rhombohedron ($OR$); their edges have the length $a_q$ and run
into fivefold symmetry directions. The CCT can uniquely be decomposed
into $PR$'s, $OR$'s and rhombic dodecahedra consisting of two $PR$ and
two $OR$.~\cite{Hen91a, Mih96a} A rhombic dodecahedron is assigned to
each $b$ bond and a $PR$ to each $c$ bond. They are called $RD$ and
$PR_c$ respectively (see Fig.~\ref{fig:Decompose}). They both connect
nodes of the CCT.

Each triangular $Y$ face is pierced by another kind of $PR$, which we
will call $PR_Y$. It is useful to divide these $PR_Y$ into two
subclasses, called $PR_Y(C)$ and $PR_Y(D)$ according to whether it
lies mostly in a $C$ or $D$ cell respectively. Each $D$ cell contains
three more $PR$'s, called $PR_D$.  Finally, each $Z$ face contains an
$OR$. This decomposition of the CCT facilitates the description of the
decoration model in the following section.

\subsection{Decoration description} \label{Initial-Section}

We now derive the initial decoration sites for the canonical cell
tiles from the structure of the $W$-phase. As our starting point we
require that our decoration model reproduces the features of the
$W$-TiZrNi structure wherever possible.

The structure of the $W$-phase corresponds to a Fibonacci 1/1
approximant.  In the canonical cell tiling description the 1/1
approximant is given by an $A_6$ tiling. The decoration of the
$A$-cell is thus completely determined by the $W$-phase. The nodes of
the $A_6$ tiling correspond to the center of the Bergman cluster in
the $W$-phase. Thus every node of the CCT will be occupied by a
two-shell Bergman cluster and the atomic configuration surrounding the
twofold ($b$) and threefold ($c$) linkages will be the same as in the
$W$-phase. This already accounts for over 80\% of the atoms in
decoration of large tilings. We then extend our decoration model to
the interior of the larger $B$, $C$ and $D$ cells using the structure
of $W$-TiZrNi as a guide.

We use the decomposition of the larger canonical cell
tiles into prolate and oblate rhombohedron as well as the rhombic
dodecahedron to describe their decoration. Since the $A_6$ tiling can
be decomposed into $PR_c$'s and $RD$'s, their atomic decoration is
determined by the atomic structure of the 1/1~approximant (see
Fig.~\ref{fig:decoration}). Simply to limit the number of free
parameters, it is assumed that the decoration sites of the $PR_Y(C)$,
$PR_Y(D)$ and $PR_D$ are the same as for the $PR_c$. As for the $OR$,
its face decoration is forced by the need to adjoin with the other
tiles and it has no space for additional interior atoms (see
Fig.~\ref{fig:decoration}). It is noted here that the site list is the
same as for the Henley-Elser decoration of Al-Mg-Zn.~\cite{Hen86a}

\begin{figure}[p]
  \caption{Decoration of (a) the rhombic dodecahedron, (b) the prolate
    rhombohedron and (c) the oblate rhombohedron.
    The right-hand pictures show the same decorations with interiors visible.}
  \label{fig:decoration}
\end{figure}

\begin{figure}[p]
  \caption{Decoration of (a) the $PR_Y(C)$, (b) the $PR_D$,
    (c) the $PR_Y(D)$ and (c) the $OR$.}
  \label{fig:decoration2}
\end{figure}

After defining the general sites of the decoration model, the
nomenclature for the different sites, necessary for a description of
the chemical decoration, is introduced. From the $W$-phase we derived
four general types of decoration sites for the Amman tiles. These are
the vertices and the mid-edge sites of the Amman tiles, the two
diagonal sites of the $PR$, and the sites inside of the $RD$.  We
denote these four kinds of sites by the letters $\alpha$, $\beta$,
$\gamma$ and $\delta$ respectively (see Fig.~\ref{fig:decoration}).
Any further differentiation of the sites is indicated by a subscript.
The physical significance is that identically labeled sites have the
same coordinates and occupations in the decoration model.

The center and the second shell icosahedron of the Bergman cluster
decorate vertices of the Amman tiles; these are denoted $\alpha_1$ and
$\alpha_2$ respectively. The sites of the first shell of the Bergman
cluster are equivalent to edge centers of the Amman tiles and are
denoted by $\beta_1$.  The $PR_c$ has six mid-edge $\beta_2$ sites on
the plane bisecting its long axis. The two atomic sites on the long
axis of the $PR_c$ which, furthermore, also decorate the faces of the
large icosahedron of the Bergman cluster, are denoted by $\gamma_1$.
The same $\gamma_1$ site also appears in the interior of the $RD$.
There are four additional sites in the interior of the $RD$ on the
plane bisecting its long axis. These four sites form a rhombus. The
sites on its short axis are denoted by $\delta_1$ and the sites on its
long axis by $\delta_2$. All of the decoration sites discussed up to
now were directly derived from the structure of the $W$-phase.

For the $B$, $C$ and $D$ cell we assume that the local structure is
similar, with the decoration of the nodes and the $b$ and $c$ bonds
the same as derived from the $W$-phase. There are additional sites
which need to be specified for the $B$, $C$ and $D$ cell. These are
the sites connected to the $PR_Y(C)$, the $PR_Y(D)$, the $PR_D$ and
the $OR$. The acute vertex of the $PR_Y$ closest to the $Y$ face, the
three mid-edges sites adjacent to this vertex and the adjacent site on
the threefold axis are not already determined by adjacent tiles. These
sites are called $\alpha_3$, $\beta_3$ and $\gamma_2$ respectively
(see Fig.~\ref{fig:decoration2}(a)).  For the $PR_Y(D)$, a number of
sites inside the $D$ cell must be specified.  Starting from the acute
vertex away from the $Y$ face, the three mid-edge sites $\beta_5$, the
three vertices $\alpha_4$, and the ring of six mid-edge sites
$\beta_6$ need to be specified. Furthermore, the site along the
diagonal that is adjacent to the acute vertex that is away from the
$Y$ face is denoted by $\gamma_3$ (see Fig.~\ref{fig:decoration2}(c)).
For the $OR$, the mid-edge sites are labeled $\beta_4$ (see
Fig.~\ref{fig:decoration2}(d)). Along the diagonal of the $PR_D$, the
site away from the node of the $D$ cell is called $\gamma_4$. The six
mid-edge sites on the plane bisecting the $PR_D$ are denoted $\beta_7$
(see Fig.~\ref{fig:decoration2}(b)).

\section{Structural Refinement} \label{Refinement-Section}

In this section the decoration model for $i$-TiZrNi is optimized using
a constrained least-squares analysis of diffraction data incorporating
results from {\it ab initio} calculations. In this approach, the
optimization of the structural model is split into two steps. In the
first step, the chemical decoration of the atomic sites is refined by
X-ray and neutron diffraction data while the site positions are kept
fixed. The atomic positions are then optimized in a second step by
{\it ab initio} relaxations of the resulting atomic decoration for
periodic tilings. These relaxed atomic positions are then used as a
constraint to the refinement of the diffraction data.

\subsection{Diffraction data}

The sample preparation and the data acquisition was performed by
E.~H.~Majzoub; details can be found in
References~[\onlinecite{Eric_Thesis, Maj00}]. The composition of the
sample is (Ti$_{41.5}$Zr$_{41.5}$Ni$_{17}$)$_{98}$Pb$_2$. The Pb was
used, since it was assumed to accumulate in the grain boundaries due
to its low solubility in the Ti-Zr-Ni phases. The regions to which the
Pb segregates are liquid during the annealing, allowing for faster
diffusion of the transition metals, thus enhancing the formation of
larger quasicrystalline grains with longer coherence length. The
samples were arc-melted from the pure materials under an argon
atmosphere and then annealed at $600^\circ$C for 7~days under vacuum.
The X-ray diffraction pattern was measured using Cu-K$_\alpha$
radiation.  Neutron powder diffraction data for the sample was
obtained at the University of Missouri Research Reactor in
collaboration with W.~B.~Yelon. The neutron wavelength was 1.7675\AA.

The uncorrected structure factors were obtained from the powder diffraction
data by performing a least-squares fit of the data to a Gaussian peak profile,
\begin{equation}
  \label{eq:PeakShape}
  I(\theta) = I_b + I_0 \frac{1}{\sqrt{2\pi}\sigma} \exp
  \left({\frac{-(\theta-\theta_0)^2} {2\sigma^2}} \right),
\end{equation}
where the peak is centered at $\theta_0$ and has a width of $\sigma$.
The integrated intensity is given by $I_0$. Any peak broadening
related to the diffractometer or the coherence length of the sample is
absorbed in the width, $\sigma$, of the Gaussian peak profile. The
data were fitted in a small interval of $\Delta \theta \approx
1^\circ\dots2^\circ$ around the center of each peak. The dependence of
the results on the choice of the size of this interval was accounted
for by the estimated errors, which reflect the range over which the
fit parameters varied when the size of the interval over which the fit
was performed was slightly changed. The background was fitted using a
single parameter, $I_b$. Out of the 31 peaks for the neutron
diffraction data, 24 peaks were indexed to the icosahedral phase and 7
were indexed to the $\delta$-Ti$_2$Ni secondary phase. For the X-ray
data, 18 structure factors were obtained. Three of the peaks belonged
to the $\delta$-Ti$_2$Ni secondary phase and 15 to the icosahedral
quasicrystal.

Overall, the number of measured structure factors is rather small.
Furthermore, since only powder diffraction data is available for the
$i$-TiZrNi quasicrystal, a few of the measured structure factors were
superpositions of crystallographically inequivalent reflections,
further exacerbating a structural analysis. In the following two
sections the initial decoration model for canonical cell tilings is
refined to the measured structure factors constraining the atomic
positions by {\it ab initio} relaxations.

\subsection{Calculating the structure factors}

A least-squares analysis for the chemical decoration of the sites of
the structure model to the X-ray diffraction and neutron scattering
data from icosahedral Ti$_{41.5}$Zr$_{41.5}$Ni$_{17}$ was performed.
As a good approximation of the quasiperiodic tiling, a cubic 8/5 CCT
approximant with 576 nodes forming 1320 $A$, 576 $BC$ and 136 $D$
cells containing 52392 atoms per unit cell and with a lattice constant
of 98\AA\ was used.~\cite{Mih93} For the 8/5 approximant, the
distribution of the complementary coordinates of the nodes has a small
variance. It is thus reasonable to assume that this tiling is a good
approximant to the hypothetical quasiperiodic canonical cell tiling.
The decorations of the 8/5~canonical cell tiling were constructed
using the \emph{Canonical Cell Tiling Package -- ``There are the
  atoms''} by Mihalkovi{\v c}.~\cite{CCTPKG}

The icosahedral structure factor, $F_\mathrm{ico}({\bf q})$, is
approximated by averaging over the amplitudes, $F_\mathrm{app}({\bf
  q_a})$, of all the Bragg peaks in the approximant that map onto a
single orbit of icosahedral peaks given by the wave vector, ${\bf q}$,
\begin{equation} \label{eq:Average_Ampl}
  F_\mathrm{ico}({\bf q}) = \frac{1}{M_\mathrm{ico}({\bf q})}
  \sum_a M_\mathrm{app}({\bf q}_a) F_\mathrm{app}({\bf q}_a).
\end{equation}
Each corresponding orbit of reflections in the approximant is
represented by a wave vector, ${\bf q}_a$, and has the multiplicity
$M_\mathrm{app}$. $M_\mathrm{ico}$ is the multiplicity of the
icosahedral orbit. The average is taken over amplitudes instead of
intensities requiring the correct choice of phases of the structure
factors. It is therefore necessary to center the approximant
structure in six dimensional space. Details of this and a discussion
of the differences between averaging over amplitudes and intensities
can be found in Appendices~\ref{App:Diffraction}
and~\ref{App:Amplitudes}.

The X-ray and neutron diffraction data were taken at room temperature.
To account for thermal motion which decreases the structure factor,
$F_\mathrm{ico}({\bf q})$, for large momentum, $q$, an isotropic
Debye-Waller factor is used,
\begin{equation}
  f_\mathrm{DW}^s(q) = \exp\left (-\frac{B_\mathrm{iso}^s}{8\pi^2}q^2\right ).
\end{equation}
Due to the small number of measured structure factors, only a single
thermal parameter, $B_\mathrm{iso}^s$ with
$s\in\{\mathrm{Ti,Zr,Ni}\}$, for each atomic species was used. The
X-ray form factors, $f_s(q)$, and the neutron scattering lengths,
$f_s$, were taken from Reference~[\onlinecite{ITC_6.1}].

Altogether the icosahedral structure factors, $F_\mathrm{ico}$, needed
for the refinement are given by
\begin{equation} \label{eq:SF_calc}
F_\mathrm{ico}({\bf q}) = \sum_{l,s} c_s(l)\; f_\mathrm{DW}^s(q)\; f_s(q)
  \sum_a \frac{M_\mathrm{app}({\bf q}_a)}{M_\mathrm{ico}({\bf q})}
  \sum_{j \in {\cal L}_l} e^{i{\bf q}_a \cdot {\bf r}_j}.
\end{equation}
The first sum is over all classes, $l$, of distinct sites of the
decoration model. The second sum is over the three different atomic
species, $s\in\{\mathrm{Ti,Zr,Ni}\}$, decorating these classes of
sites with concentrations, $c_s(l)$. The third summation is over the
orbits of the approximant structure represented by the wave vector,
${\bf q}_a$, which correspond to the icosahedral wave vector, ${\bf
  q}$.  Finally the last summation is taken over the individual atomic
site, ${\bf r}_j$, belonging to the set of sites, ${\cal L}_l$, of the
site class, $l$.

\subsection{Constrained least-squares refinement}

Before we can compare the experimental peak intensities to the
calculated structure factors, several correction factors have to be
taken into account. For both, the X-ray and the neutron diffraction
data, the intensities must be corrected by the Lorentz factor,
$\sec\theta\; \mathrm{cosec}^2\theta$. Furthermore for the X-rays, the
intensities are modified by the polarization, $P$, of the beam as
described by the Thomson formula
\begin{equation} \label{eq:Thompson}
  \frac{1+P \cos^2(2\theta)}{1+P}.
\end{equation}
Since the X-ray beam was unpolarized, $P=1$ is used. Overall the
estimated intensity of a reflection is given by
\begin{equation} \label{eq:Int_x}
  I_\mathrm{ico}(q) = \frac{1+P\cos^2(2\theta)}{(1+P)(\cos\theta \sin^2\theta)}
  \; \left | F_\mathrm{ico}(q) \right |^2
\end{equation}
for X-ray diffraction and by
\begin{equation} \label{eq:Int_n}
  I_\mathrm{ico}(q) = \sec\theta\; \mathrm{cosec}^2\theta \; \left
    | F_\mathrm{ico}(q) \right |^2
\end{equation}
for neutron scattering.

Since the positions of the atoms are kept fixed in the refinement, the
last two summations in Equation~(\ref{eq:SF_calc}) over the atomic
sites of each class with identical atomic decoration and over the
orbits of wave vectors for the approximant are performed beforehand
for all necessary values of ${\bf q}_\mathrm{ico}$. The total
structure factor for each ${\bf q}$ was then obtained by summing over
these partial structure factors multiplied by the average form factor
or scattering length, $f_s(q)$, and the average Debye-Waller factor of
the atoms decorating the site. The corrected intensity,
$I_\mathrm{ico}$, is then obtained from the calculated structure
factor, $F_\mathrm{ico}$, using Equations~\ref{eq:Int_x}
or~\ref{eq:Int_n} respectively.

From the calculated intensities, the residuals $R$ and $R_w$ as well
as the reduced $\chi^2$ are obtained, providing a standard measure of
how well the calculated data matches the observed one.  The residuals
are defined as
\begin{equation}
  R = \frac{\sum_q \left | \sqrt{I_\mathrm{exp}(q)} - \sqrt{I_\mathrm{ico}(q)}
  \right |}{\sum_q \sqrt{I_\mathrm{exp}(q)}}
\end{equation}
and
\begin{equation}
  R_w = \sqrt{\frac{\sum_q w(q) \left (\sqrt{I_\mathrm{exp}(q)} -
  \sqrt{I_\mathrm{ico}(q)}\right)^2}{\sum_q w(q) I_\mathrm{exp}(q)}}
\end{equation}
where the sums are over all observed and calculated intensities,
$I_\mathrm{exp}$ and $I_\mathrm{ico}$ respectively. The weights $w$
are given by the uncertainty of the measured intensities where
$w=1/\sigma_I^2$.  The ``goodness'' of the fit is given by the reduced
$\chi^2$,
\begin{equation}
  \chi^2 = \frac{\sum_q w(q) \left (\sqrt{I_\mathrm{exp}(q)} - \sqrt{I_\mathrm{ico}(q)}
  \right)^2}{N-N_p},
\end{equation}
where $N$ is the total number of reflections and $N_p$ is the number of
estimated parameters.

Since the number of measured reflections is rather small, additional
measured quantities such as the composition of the structure can be
incorporated as constraints to the refinement and to limit the
parameter space. The composition of the refined structure, $c_i$, was
constrained by a penalty function,
\begin{equation}
  \label{eq:penalty}
  \sum_{i \in\{\mathrm{Ti, Zr, Ni}\}}\left ( c_i -c_i^o \right )^2/\sigma_c^2.
\end{equation}
In order to ensure that the composition of the refined structure,
$c_i$, does not deviate by more than a few percent from the nominal
experimental value, $c_i^o$, a value of $\sigma_c=0.02$ was used.
The $\chi^2$ was then minimized by varying the chemical decoration of
the sites using Powell's conjugate direction method.~\cite{NumRec}

The chemical decoration of the ideal atomic sites is refined with
respect to the structure factors of 38 reflections for $i$-TiZrNi, 15
X-ray and 23 neutron diffraction structure factors. The maximum number
of different sites of the decoration model is 18. The decoration of
the sites in the $W$-phase structure was used for the initial
configuration, i.e. $\alpha$ sites were decorated with Ni, $\beta$
sites with Ti and $\gamma$ and $\delta$ sites with Zr. The chemical
decoration on each of these site classes was allowed to vary between
two components, either Ti and Zr or Ti and Ni. About a dozen different
starting configurations were used. The majority of the optimization
runs resulted in the same final concentrations and gave the lowest
observed $\chi^2$. However, as common in optimization problems, we
cannot guarantee to have found the global minimum. During the
refinement classes of sites which showed similar chemical occupancies
were combined into a single class. The final number of independent
classes of sites was six. This, and the three thermal parameters yield
a total of nine estimated parameters.

The refinement using the initial decoration sites yielded for the
residuals, $R$, of the structure factors 11.5\% for X-ray, 6.1\% for
neutron and 6.9\% combined. The weighted residuals, $R_w$, are 15.7\%
for X-ray, 4.6\% for neutron and 5.7\% combined. The reduced $\chi^2$
is 1.8 indicating crystallographic significance for the fit. The
thermal parameters for Ti and Zr, $B_\mathrm{iso}^\mathrm{Ti} =
1.1$\AA$^2$ and $B_\mathrm{iso}^\mathrm{Zr} = 1.2$\AA$^2$, are
comparable to values found in the $\delta$-(Ti,Zr)$_2$Ni
alloys.~\cite{Mackay94} The thermal parameter for Ni,
$B_\mathrm{iso}^\mathrm{Ni} = 2.3$\AA$^2$, is about a factor two
larger than the value found in $\delta$-(Ti,Zr)$_2$Ni. This could
reflect static disorder on some of the Ni sites.

\subsection{{\it Ab initio} relaxations}

So far we have only refined the chemical decoration of the ideal sites
of the canonical cell tiling. In the next step we perform {\it ab
  initio} relaxations for the resulting structure to investigate the
site positions.  The {\it ab initio} calculations were performed with
VASP~\cite{Kresse96a,Kresse96b,Kresse93}, which is a density
functional code using a plane-wave basis and ultrasoft Vanderbilt-type
pseudopotentials~\cite{Vanderbilt90,Kresse94}.  Atomic-level forces
are calculated and relaxations with a conjugate gradient method are
performed.  {\it Ab initio} calculations are computationally very
expensive. Thus, only calculations for small systems with up to a few
hundred atoms are feasible on current computers. Therefore, {\it ab
  initio} relaxations only for the small periodic tilings of the
canonical cells were performed. The tilings used were the $A_6$, the
$B_2C_2$ and the $D_2$ tiling.~\cite{Hen91a}

As a starting configuration, the results from the analysis of the
diffraction data were used. The chemical order was idealized by
placing only the majority species on each site. The calculations were
performed using the generalized gradient approximation by Perdew and
Wang~\cite{Perdew92a}, and a plane-wave kinetic-energy cutoff of
$E_{cut} = 302.0\,$eV was chosen to ensure convergence. The
pseudopotentials for Ti and Zr describe, besides the usual outer shell
states, also the 3p and 4p states, respectively, as valence states. Due
to the size of the quasicrystal-like tiling structures a check of the
convergence with the number of k-points was not always possible due to
computational limitation. Therefore the k-point mesh was chosen as
large as possible, i.e.  $3 \times 3 \times 3$ for the $A_6$ tiling,
$2 \times 2 \times 2$ for the $B_2C_2$ tiling and $1 \times 1 \times
1$ for the $D_2$ tiling, corresponding to 4, 2 and 1 ${\bf k}$-points
in the irreducible part of the Brillouin zone respectively. The
positions of the atoms, as well as the shape and volume of the unit
cells, were relaxed until the total electronic energy change was
smaller than $1\,$meV. This corresponds to atomic-level forces
$F_{max} \leq 0.02\,$eV/\AA.

\begin{table}[htbp]
  \caption{Maximum and average displacements in [\AA] of the atoms from their
    ideal positions for the different periodic canonical cell tilings
    after relaxation.}
  \label{tab:DisplacementsIdeal}
  \begin{center}
    \begin{tabular}{c c c c}
      \toprule
      Tiling & Atoms & \multicolumn{2}{c}{Displacement} \\
      & & average (rms) & max \\
      \colrule
      $A_6$    &  81 & 0.10 & 0.19 \\
      $B_2C_2$ &  91 & 0.11 & 0.22 \\
      $D_2$    & 123 & 0.13 & 0.25 \\
      \botrule
    \end{tabular}
  \end{center}
\end{table}

The atomic positions after the {\it ab initio} relaxation were
remarkably close for all three tilings to the ideal atomic positions,
such as the corners, mid-edges etc.\ of the decorated tiles, with
average displacements smaller than 0.13\,\AA\ (see Table
\ref{tab:DisplacementsIdeal}).

The resulting atomic positions from the relaxation calculations were
then symmetrized. This involves mapping the relaxed positions back
onto their canonical orbit and calculating the average canonical
coordinates of equivalent atoms.~\cite{CCTPKG,Mih96b} The deviations
of the relaxed atomic positions from their canonical coordinates are
in general rather small with an average deviation of $0.07\,$\AA.  The
largest deviation occurs for the $\beta_4$ site, where the relaxed
positions deviates by $0.19\,$\AA\ in the $D_2$ and by $0.05\,$\AA\ in
the $B_2C_2$ cell. In fact, the deviations in both cells have almost
opposite directions.

The symmetrized positions are used as an input for fitting the site
chemistry to the diffraction data. This process can, in principle, be
repeated until convergence is achieved. In our case, since no
significant changes in the decoration were observed, the optimization
was already converged after the refinement of the relaxed atomic
positions.

\begin{figure}[htbp]
  \caption{Comparison of the experimental X-ray diffraction and neutron
    scattering data to the simulated diffraction of the icosahedral
    structure.}
  \label{fig:Diffraction}
\end{figure}

\begin{table}[tbp]
  \caption{Results of the refinement of the chemical decoration for the
      quasicrystal using the relaxed atomic positions.}
  \label{tab:DecorationResults}
  \begin{center}
    \begin{tabular}{l l l l c c}
      \toprule
      \multicolumn{3}{l}{Site} & Description & Decoration & Class \\
      \colrule
      $\alpha_1$ & Cluster && Center               & Ni (1.0) & 1 \\ 
      $\beta_1$  &         && Small icosahedron    & Ti (1.0) & 2 \\ 
      $\alpha_2$ &         && Large icosahedron    & Ni (1.0) & 1 \\ 
      $\gamma_1$ &         && Faces                & Zr (1.0) & 3 \\ 
      \colrule
      $\delta_1$ & $b$-bond & $RD$   & Close pair   & Zr (1.0) & 3 \\ 
      $\delta_2$ &          & $RD$   & Distant pair & Zr (1.0) & 3 \\ 
      \colrule                  
      $\beta_2$  & $c$-bond & $PR_c$ & Mid edge     &
      $\left \{\! \begin{array}{c} \mathrm{Ti~(0.8)} \\ \mathrm{Zr~(0.2)} \end{array} \right .$
      & 4 \\ 
      \colrule                  
      $\alpha_3$ & $Y$-face &$PR_Y(C)$& Vertex       & Ni (1.0) & 1 \\ 
      $\beta_3$  &          &$PR_Y(C)$& Mid-edge     & Ti (1.0) & 2 \\ 
      $\gamma_2$ &          &$PR_Y(C)$& Upper diagonal &
      $\left \{\! \begin{array}{c} \mathrm{Ti~(0.6)} \\ \mathrm{Zr~(0.4)} \end{array} \right .$
      & 5 \\ 
      \colrule
      $\beta_4$  & $Z$-face & $OR$ & Mid-edges    &
      $\left \{\! \begin{array}{c} \mathrm{Ni~(0.7)} \\ \mathrm{Ti~(0.3)} \end{array} \right .$
      & 6 \\ 
      \colrule
      $\gamma_3$ & $D$-cell &$PR_Y(D)$&Lower diagonal& Zr (1.0) & 3 \\ 
      $\alpha_4$ &          &$PR_Y(D)$& Vertices    & Ni (1.0) & 1 \\ 
      $\beta_7$  &          &$PR_Y(D)$& Mid-edges   & Ti (1.0) & 2 \\ 
      $\gamma_4$ &          & $PR_D$ & Diagonal     & Ti (1.0) & 2 \\ 
      $\beta_5$  &          & $PR_D$ & Mid-edges    & Ti (1.0) & 2 \\ 
      $\beta_6$  &          & $PR_D$ & Mid-edges    & Ti (1.0) & 2 \\ 
      \botrule
    \end{tabular}
  \end{center}
\end{table}

Using the relaxed positions, the residuals, $R$, of the structure
factors decrease to 10.8\% for X-ray, 5.1\% for neutron and 5.9\%
combined. The weighted residuals, $R_w$, are 15.0\% for X-ray, 4.0\%
for neutron and 5.1\% combined. The reduced $\chi^2$ is 1.4,
indicating a good agreement between the calculated and measured
structure factors (see Fig.~\ref{fig:Diffraction}). Overall, the
relaxed positions lead to a better agreement with the diffraction data
than the unrelaxed positions.

The resulting chemical decoration is given in Table
\ref{tab:DecorationResults}. The thermal parameter decreases for Ti to
0.3\AA$^2$, for Zr to 1.0\AA$^2$ and it is unchanged for Ni. The
chemical decorations on the sites change by less than 5\%, with the
exception of site $\gamma_2$, where the probability of finding Ti
increases by 10\%, and the site $\alpha_4$, that is now only occupied
by Ni. The stoichiometry changes by less than 0.5\%. The composition
of the refined structure Ti$_{41}$Zr$_{40}$Ni$_{19}$ agrees with the
nominal composition Ti$_{41.5}$Zr$_{41.5}$Ni$_{17}$ to within 2\%.

This is the first refinement that systematically combines diffraction
data and structural energies. Since the powder diffraction data for
$i$-TiZrNi are not available for sufficiently high momentum $q$ to
usefully constrain the coordinates the atomic positions were fixed to
the results from {\it ab inito} relaxations. If better data were
available for this quasicrystal, positional information would be
available from the diffraction data as well. The relaxed atomic
positions lead to an improved agreement between the calculated and
measured structure factors. This illustrates that results from {\it ab
  initio} relaxations can successfully be incorporated as constraints
into structural refinements.

\section{Structural features of the quasicrystal}
\label{Sec:Structure}

In this section we discuss some details of the resulting structural
model for $i$-TiZrNi using the large 8/5 approximant and compare to
the structure of the $W$-phase. Results for pair correlation
functions, coordination numbers and local atomic environments, which
are important for comparisons to experimental data such as EXAFS, are
given in Tables~\ref{tab:W-Neighbours} and~\ref{tab:85-Neighbours} as
well as Figure~\ref{fig:85-pdf-partial}.

\begin{figure}[p]
  \caption{Partial pair distribution functions of the $W$-phase (top) and 
  the 8/5-approximant model (bottom).}
  \label{fig:85-pdf-partial}
\end{figure}

\begin{table}[htbp]
  \caption{Number and average distance of nearest neighbors in parentheses in
    [\AA] in the $W$-phase (a=14.3\AA).}
  \label{tab:W-Neighbours}
  \begin{center}
    \renewcommand{\tabcolsep}{0.1cm}
    \begin{tabular}{c c c c c}
      \toprule
      Center & Ti & Zr & Ni & Total \\
      \colrule
      Ti & 4.91 (2.90) & 5.12 (3.07) & 2.00 (2.66) & 12.03 (2.93)\\
      Zr & 7.97 (3.07) & 1.76 (3.12) & 2.75 (3.19) & 12.48 (3.10)\\
      Ni & 6.37 (2.66) & 5.63 (3.19) & 0.00 (-)        & 12.00 (2.91)\\
      \botrule
    \end{tabular}
  \end{center}
\end{table}
\begin{table}[htbp]
  \caption{Number and average distance of nearest neighbors in parentheses in
    [\AA] in the 8/5 approximant model (a=98.0\AA).}
  \label{tab:85-Neighbours}
  \begin{center}
    \renewcommand{\tabcolsep}{0.1cm}
    \begin{tabular}{c c c c c}
      \toprule
      Center & Ti & Zr & Ni & Total \\
      \colrule
      Ti & 3.95 (2.88) & 5.85 (3.04) & 2.34 (2.67) & 12.14 (2.92)\\
      Zr & 5.91 (3.04) & 3.30 (3.02) & 2.91 (3.13) & 12.12 (3.06)\\
      Ni & 4.93 (2.67) & 6.05 (3.13) & 0.83 (2.70) & 11.81 (2.91)\\
      \botrule
    \end{tabular}
  \end{center}
\end{table}

Figure~\ref{fig:85-pdf-partial} shows the partial pair-distribution
functions for the 1/1 approximant phase, $W$-TiZrNi, and the
8/5-approximant. Tables~\ref{tab:W-Neighbours}
and~\ref{tab:85-Neighbours} list the numbers and average distances of
nearest neighbors in both structures. The pair-correlation functions
for the $W$-phase and the quasicrystal resemble another, corresponding
to the similar local atomic structures of both phases. While the
$W$-phase has no Ni-Ni nearest neighbors at all, the number of Ni-Ni
neighbor pairs in the quasicrystal is rather small. Chemical disorder
was found only on a small number of sites, apparently localized to the
glue region between the Bergman cluster, similar to the $W$-TiZrNi
approximant (see Tab.~\ref{tab:Approximant}). Results of recent
investigations of the local atomic structure of the $W$-phase and the
quasicrystal by EXAFS measurements~\cite{Sadoc00} are consistent with
the pair-distribution functions of the structural model developed
here.

The features of the decoration model of the quasicrystal can be
understood by the energetics of the interactions between the
constituents. Ti and Zr atoms are completely miscible and exhibit a
zero heat of mixing. Ti and Ni as well as Zr and Ni, on the other
hand, have a strongly negative heat of mixing, indicating strong
attractive interactions between these pairs. Thus, it is energetically
favorable for all the Ni atoms to be surrounded by Ti or Zr. This
explains why, in the structure of the 1/1 approximant and the
quasicrystal, hardly any Ni-Ni bonds occur. Since Zr is slightly
larger than Ti it is no surprise that Zr occupies the more open sites
of the structure given by the interior of the $PR$ and $RD$.

The refined structure of the quasicrystal is very similar to the
structure of the 1/1 approximant phase, $W$-TiZrNi. The atomic sites
for the structure model of the quasicrystal are based on the $W$-phase
structure. The refined decoration of these sites is the same in most
cases. However, the $\delta_1$ and $\delta_2$ sites along the $b$ bond
are occupied by Zr only in the quasicrystal, while in the $W$-phase
chemical disorder occurs between Ti and Zr. Overall, the amount of
chemical disorder in the quasicrystal appears to be of the same order
or even slightly less than in the $W$-phase.  Considering the apparent
disorder in the powder diffraction patterns, the lack of chemical
disorder on most sites of the structural model of the quasicrystal
seems surprising at first and might be fortuitous. However, it is
noted, that the quasicrystal forms at low temperatures, possibly
reducing the amount of entropic disorder.  We speculate, that an
(aperiodic) $i$-TiZrNi quasicrystal, or a very similar large-cell
crystal, could be a ground state phase. This idea is investigated in a
subsequent paper.

\section{Conclusion}

We have developed the first realistic structural model for the
icosahedral Ti-Zr-Ni quasicrystal. Our model is based on atomic
decorations of the canonical cell tilings. For the structural
refinement a new method of constrained least squares analysis is used,
combining X-ray and neutron diffraction data with results from {\it ab
  initio} relaxations. The refined structural model exhibits only
small amounts of chemical disorder between Ti and Zr.

This structural model should be helpful for a better understanding and
interpretation of experimental data such as, for example, EXAFS data,
pcT-isotherms, relations to crystalline phases and the ternary phase
diagram.  Furthermore this model will enable us to investigate the
possibility of $i$-TiZrNi as a ground state quasicrystal.

\begin{acknowledgments}
  The work at Washington University was supported by the National
  Science Foundation (NSF) under grants DMR 97-05202 and DMR 00-72787.
  Work at Cornell made use of the Cornell Center for Materials
  Research Computing Facilities, supported through the National
  Science Foundation MRSEC program (DMR-0079992).  C.L.H.\ was
  supported by the Department of Energy (DOE) grant
  DE-FG02-89ER-45405.  R.G.H.\ was partially supported by NSF grant
  DMR-0080766 and by DOE grant DE-FG02-99ER45795. We thank Alex Quandt
  and Jeff Neaton for discussions about the VASP code, Marek
  Mihalkovi{\v c} for the code to decorate canonical cell tilings and
  E. H.  Majzoub for providing us with the X-ray and neutron
  diffraction data.
\end{acknowledgments}

\begin{appendix}
\section{Diffraction from Approximants}
\label{App:Diffraction}

In this Appendix a method is presented for the calculation of the
structure factors of the icosahedral quasicrystal from large cubic
approximants.

Assuming that the local structure of the approximant closely resembles
the structure of the quasicrystals, the quasicrystal structure factors
can be approximated by the structure factors of a large periodic
approximant.  However, a direct comparison of the structure factors of
the approximant to measured structure factors of the icosahedral phase
is not possible due to the different symmetries of the structures.
This leads to different multiplicities of corresponding reflections
for both structures and a splitting of orbits of icosahedral
reflections in the case of the approximant. The problem is further
complicated by the fact that the Bragg reflections of the quasicrystal
form a dense set, which is not the case for the periodic approximants.
Inequivalent low intensity reflections, therefore, can become
degenerate in the approximant. The problem of degenerate reflections
in the approximant becomes increasingly important in smaller
approximants. Since, for structural investigations only Bragg
reflections with intensities above a threshold are used, the
degeneracy problem can be avoided by using a large enough approximant.

The six dimensional representation of icosahedral quasicrystals
enables one to relate their diffraction patterns to the diffraction
patterns of their related crystalline approximants. The reciprocal
lattice (RL) of the approximant is obtained when the golden mean,
$\tau$, which appears in the icosahedral basis of the physical space,
is replaced by the rational number $\tau_n=f_{n+1}/f_n$. The mapping
of the icosahedral RL to the cubic RL is induced by keeping the same
six indices but using the rationally related basis vectors. Every
icosahedral reciprocal lattice vector (RLV) maps to a unique cubic
RLV. The maximum cubic subgroup of the icosahedral symmetry group,
$m\bar 3\bar5$, is given by $m3$.  While the order of the icosahedral
group is 120, the order of the cubic subgroup is only 24. Thus, the
image of one orbit of equivalent icosahedral reflections is the union
of up to five cubic orbits. Although an icosahedral RLV vector does
not map to a cubic RLV in the same direction, any icosahedral RLV in a
given fundamental domain of the cubic group maps to a cubic RLV in the
same fundamental domain.  Thus, to get the standard representative
cubic RLV's corresponding to one icosahedral RLV orbit, we only need
to check the images of all icosahedral RLV's in that orbit whose
orientations fall in the standard principal triangle of the cubic
group.

The calculation of the structure factors for the icosahedral phase
from the cubic approximant proceeds in two steps. First, for every
orbit of icosahedral reflections with complementary components smaller
than some cutoff, $Q^\perp_\mathrm{max}$, the multiplicity,
$M_\mathrm{ico}$, is determined by analyzing if the representing RLV
points into any high-symmetry direction of the icosahedral point
group. The icosahedral RLV's of the orbit which fall into the
fundamental domain of the cubic point group are determined.  Depending
on the multiplicity of the icosahedral orbit this results in a set of
up to five RLV's. For each of these icosahedral RLV, the corresponding
cubic RLV in the approximant is determined by the projection method,
where the golden mean, $\tau$, in the basis vectors of the physical
space is replaced by the rational approximant, $\tau_n$. The
multiplicity, $M_\mathrm{a}$, of the orbit represented by each of the
cubic RLV's is calculated by analyzing if the cubic RLV points into
any high-symmetry direction of the cubic point group.

In the second step, to approximate the structure factor of the
icosahedral RLV, the structure factors of the corresponding cubic
RLV's of the periodic approximant are calculated and averaged. This
average could be taken over either the amplitudes or the intensities
of the structure factors. While for reflections with a high intensity,
both methods of averaging result in nearly the same values, for low
intensity reflections the right choice of averaging is more important.
Since all the cubic approximants are centrosymmetric, the structures
can be centered such that the phases of the structure factors are
either 0 or $\pi$, resulting in a plus or minus sign for the structure
factors.  The correct centering of the approximant structures involves
a shift in the six-dimensional space.  The 1/1 approximant structure
requires no shift. For the cubic 2/1, 3/2, 8/5 and 13/8 approximants
of the canonical cell tiling, the six-dimensional centering vector is
given by $\frac{1}{2} (1,0,1,1,0,0)$. These approximants have the
symmetry group $Pa\bar 3$.  That means that there is one 3-fold axis
which passes through the node at the origin in real space. That 3-fold
axis is chosen to be in the cubic $(111)$ direction which corresponds
to the six-dimensional vector $(1,0,1,1,0,0)$.

For reflections with nearly vanishing intensity, the splitting of the
reflection in the approximant leads to a distribution of structure
factors around zero. An averaging over intensities would lead to a
positive value, while for averaging over amplitudes the resulting
value is closer to zero and provides a better approximation. In this
work, the icosahedral structure factor is estimated by averaging over
the amplitudes of the corresponding structure factors of a large cubic
approximant (see Eqs.~(\ref{eq:Average_Ampl}) and~(\ref{eq:SF_calc})).
In order for the approximated structure factors to closely resemble
the icosahedral structure factors, the atomic structure of the
approximant has to resemble the structure of the icosahedral phase
closely, which is more likely the case for larger approximants.

\section{Averaging over Amplitudes or Intensities} \label{App:Amplitudes}

To compare the accuracy of the prediction of the structure factors
from averaging over amplitudes and over intensities the method is
applied to the Penrose tiling, since for the ideal quasiperiodic
Penrose tiling, the structure factors can be easily calculated by
Fourier transformation of the acceptance domain. If the vertices of
the Penrose tiling are identical scatterers, the structure factors of
the ideal quasiperiodic Penrose tiling are determined by the Fourier
transformation of the triacontahedral acceptance
domain.~\cite{Elser86}

\begin{figure}[p]
  \caption{Comparison of the structure factors of the
    icosahedral Penrose tiling to the estimated values from the
    periodic 1/1 and 3/2 approximants. The results from averaging
    over intensities is show in the top panels and for averaging
    over amplitudes in the bottom panels.}
  \label{fig:SF_Penrose}
\end{figure}

In Figure~\ref{fig:SF_Penrose} the structure
factors for the ideal quasiperiodic Penrose tiling are compared to the
estimated values from the cubic Penrose 1/1 and 3/2 approximant
structures. While for the 1/1 approximant the structure factors
deviate strongly at lower intensities, already for the 3/2
approximants the estimated structure factors closely resemble the
icosahedral structure factors over about three orders of magnitude.
Averaging over amplitudes instead of intensities leads to significant
improvements in the case of the 1/1 approximant, but only minor
changes for the 3/2 approximant. It is noted here, that the Penrose
tiling decorated with unit scatterers on the vertices is likely to
oversimplify the problem and that in more realistically decorated
structures, larger deviations could occur.

\end{appendix}

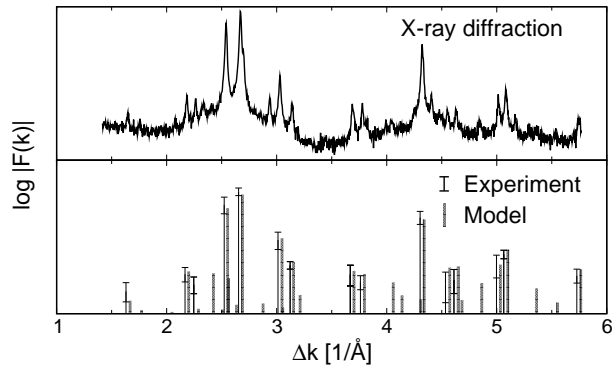
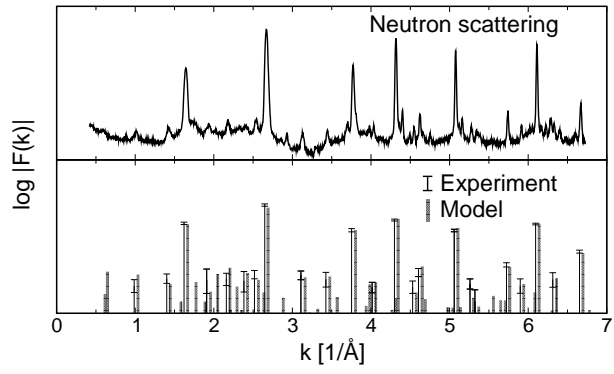
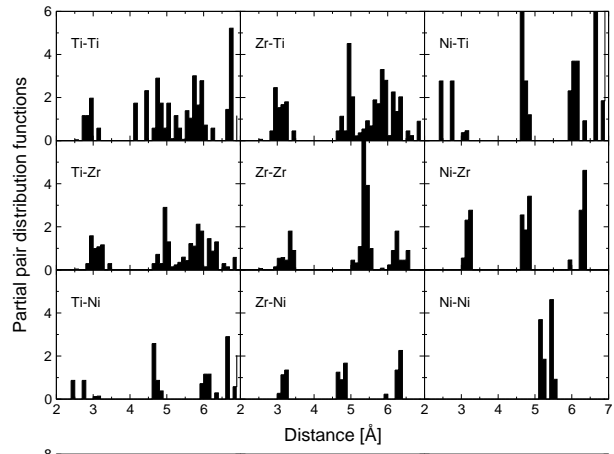
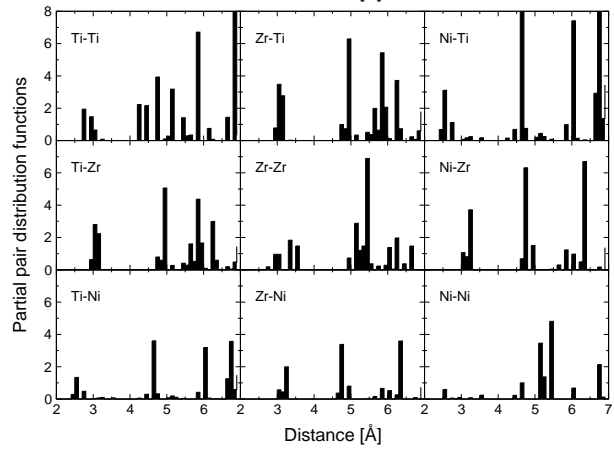
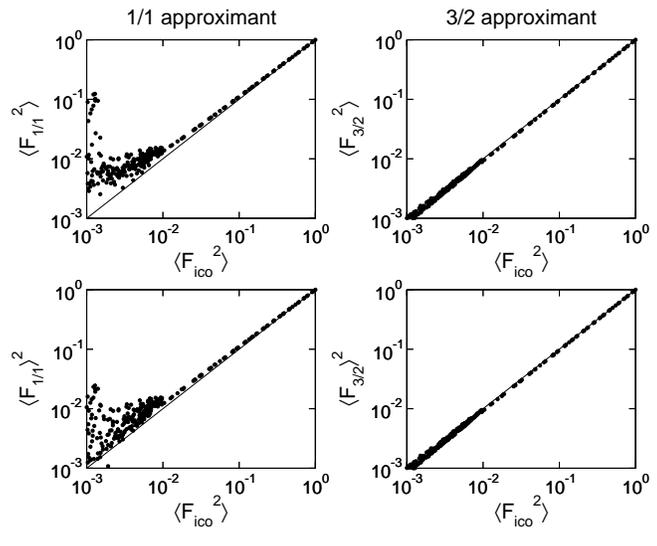
\end{document}